\documentclass[aps,prd,onecolumn,10pt]{revtex4-1}
\usepackage{graphicx}
\usepackage{hyperref}

\newcommand{\be}{\begin{equation}}
\newcommand{\ee}{\end{equation}}
\newcommand{\bea}{\begin{eqnarray}}
\newcommand{\eea}{\end{eqnarray}}

\begin{document}
\title{Next-to-leading order gravitational spin1-spin2 coupling\\ with Kaluza-Klein reduction}
\author{Michele Levi}
\email{michele@phys.huji.ac.il}
\affiliation{Racah Institute of Physics, Hebrew University, Jerusalem 91904, Israel}
\date{\today}

\begin{abstract}
We use the recently proposed Kaluza-Klein (KK) reduction over the time dimension, within an effective field theory (EFT) approach, to calculate the next-to-leading order gravitational spin1-spin2 interaction between two spinning compact objects. It is shown here that to next-to-leading order in the spin1-spin2 interaction, the reduced KK action within the stationary approximation is sufficient to describe the gravitational interaction, and that it simplifies calculation substantially. We also find here that the gravitomagnetic vector field defined within the KK decomposition of the metric mostly dominates the mediation of the interaction. Our results coincide with those calculated in the Arnowitt-Deser-Misner Hamiltonian formalism, and we provide another explanation for the discrepancy with the result previously derived within the EFT approach, thus demonstrating clearly the equivalence of the Arnowitt-Deser-Misner Hamiltonian formalism and the EFT action approach.  
\end{abstract}

\maketitle	

\section{Introduction}

A novel effective field theory (EFT) approach treating the post-Newtonian (PN) formalism of general relativity (GR) was introduced recently by Goldberger and Rothstein \cite{Goldberger:2004jt}. It is very advantageous in applying the efficient standard tools of quantum field theory to GR, notably handling the regularization required for higher order corrections in the PN approximation with the standard renormalization scheme. Moreover, an EFT approach enables us to handle a variety of physical situations, which are characterized by having multiple typical length scales. Initially, it was implemented on the evolution of an inspiralling binary to yield predictions of gravitational radiation \cite{Goldberger:2004jt, Goldberger:2007hy}. Later, it was used to obtain higher order thermodynamic results for higher dimensional Kaluza-Klein (KK) black holes \cite{Chu:2006ce}. Recently, it was further improved, and used to obtain first thermodynamic properties of higher dimensional KK \textit{rotating} black holes \cite{Kol:2007rx}. 

In this paper, we make a first application in the PN approximation of the KK reduction over the time dimension, proposed recently in \cite{Kol:2007bc} within the EFT approach, to obtain the next-to-leading order (NLO) spin1-spin2 interaction of a binary of spinning compact objects. The KK reduced action in the stationary approximation is shown here to be sufficient to describe the gravitational interaction to NLO in the spin1-spin2 sector. It is also shown to greatly simplify calculation, as well as provide further physical insight on the mediation of the interaction with spin. Our results coincide with those calculated in the Arnowitt-Deser-Misner (ADM) Hamiltonian formalism \cite{Steinhoff:2007mb}, demonstrating clearly its equivalence with the EFT action approach. Here, we provide another explanation for the discrepancy with the result previously derived within the EFT approach in \cite{Porto:2006bt}, which with the addition of \cite{Porto:2007tt}, was shown in the latter to be canonically related to the ADM result. 

After the completion of this work, \cite{Porto:2008tb} appeared which partially overlaps it.

\section{Spinning objects in an EFT approach with Kaluza-Klein reduction}

To the order that we are calculating here, the action describing the dynamics of two spinning objects is given by 
\be
S = -\sum_{a=1}^2 \left (\int m_a d\tau_a + \frac{1}{2}\int S_a^{\mu\nu}\Omega^a_{\mu\nu}(\lambda_a) d\lambda_a \right) - \frac{1}{16\pi G} \int d^4x \left(\sqrt{g} R+ \cal{L}_{\it{GF}}\right),
\ee
where $\tau_a$ and $\lambda_a$ are the proper time and worldline parameters of the $a$th particle's worldline \cite{Porto:2005ac}. The first term is the relativistic point particles (PP) interacting with gravity, where the PP spin coupling appears as the second term here. $\Omega_{\mu\nu}$ is the particle generalized angular velocity, expanded around flat spacetime in terms of tetrads and the connection of the metric, with the spin variable $S^{\mu\nu}$ conjugate to it. The last part of the action is the gravitational field interaction, including the usual Einstein-Hilbert (EH) action together with a gauge fixing term. Other higher order terms in the action do not contribute to the calculation of the NLO spin1-spin2 interaction as explained in \cite{Levi:2010zu}. The conventions used here are $c=1$, a $(+,-,-,-)$ signature, and $R^\mu_{~\nu\alpha\beta}=\partial_\alpha\Gamma^\mu_{\nu\beta}-\partial_\beta\Gamma^\mu_{\nu\alpha}+\cdots$. 

In the expansion of the metric around flat spacetime the metric is decomposed into potential and radiation modes, as explained in \cite{Goldberger:2004jt}. Both modes have the same typical time variation scale set by the binary orbital frequency $v/r$, but there is a different typical length scale for each, $r$ and $r/v$, respectively, where $r$ and $v$ are the typical orbital parameters of the binary, and we are working in the PN limit $v\ll1$. Thus, as far as the potential gravitons are concerned, the radiation component is just a slowly varying background field of soft momenta gravitons. Moreover, considering these typical scales, we see that the potential gravitons are off shell, with their frequency being much smaller than their momentum, and thus can be approximated as stationary to leading order. 

Based on these observations a KK reduction over the time dimension of the potential field was suggested and used to simplify the gravitational action, and consequently the EFT calculation \cite{Kol:2007rx,Kol:2007bc}. At first stage, the metric is parametrized according to the Kaluza-Klein ansatz
\be
ds^2 = e^{2 \phi}(dt - A_i\, dx^i)^2 -e^{-2 \phi}\, \gamma_{ij}\,dx^i dx^j.
\ee
This just defines a set of new fields $(\phi, A_i, \gamma_{ij}), ~i,j=1,2,3$, the scalar field $\phi$, which corresponds to the Newtonian potential, the gravitomagnetic vector field $A_i$, and the 3-dimensional symmetric 2-tensor field $\gamma_{ij}$, the nonrelativistic gravitational (NRG) fields as discussed in \cite{Kol:2007bc}. Next, we suppress the time dependence of the fields to obtain the KK reduced action for the gravitational field, given by
\be
S_{KK} = -\frac{1}{16\pi G} \int dt\, d^3x \sqrt{\gamma} \left[ -R[\gamma_{ij}] + 2\,\gamma^{ij}\, \partial_i \phi\,\partial_j \phi\,  -  \frac{1}{4}\, e^{4\phi} F_{ij}F_{kl}\gamma^{ik}\gamma^{jl}\right],  
\ee
where $F_{ij}\equiv\partial_i A_j-\partial_j A_i$. After obtaining the KK reduced action the gauge fixing term is to be chosen to set completely the gravitational action.

Thus, it remains to define the different NRG graviton propagators, which are used in our calculations. The propagators can be read off from the terms that are quadratic in the fields in the gravitational action, depending on the choice of the gauge fixing term of course. We use here the natural gauge adequate for the NRG fields, namely, the Lorentz gauge for the vector field $A_i$, and the linear harmonic gauge for the 2-tensor field $\gamma_{ij}$ in 3D. Thus, the gauge fixing term reads
\be\label{eq:gf}
S_{\it{GF}} = \frac{1}{32\pi G} \int dtd^3x\left[\left(\partial_{i}A_i\right)^2 - \left(\partial_j\gamma_{ij}-\frac{1}{2}\partial_i\gamma_{jj}\right)^2\right].
\ee
Quadratic terms with time derivatives, such as what we shall see in Eq.~(\ref{eq:2gc}), are suppressed as subleading corrections in powers of $v^2$ as explained above: the potential graviton propagators are instantaneous within the leading stationary approximation, representing off shell gravitons. The NRG scalar, vector, and tensor graviton propagators in the harmonic gauge are thus given by 
\be \label{eq:props}
\begin{array}{rcrcll}
\langle {\phi(\bf{x_1})}{\phi(\bf{x_2})} \rangle &=& \frac{1}{8}&\int \frac {d^3{\bf k}}{(2\pi)^3}{1\over {\bf k}^2} e^{i{\bf k}\cdot\left({\bf x}_1 - {\bf x}_2\right)},&&\\
\langle {A_i(\bf{x_1})}{A_j(\bf{x_2})} \rangle &=& -\frac{1}{2}&\int \frac {d^3{\bf k}}{(2\pi)^3}{1\over {\bf k}^2} e^{i{\bf k}\cdot\left({\bf x}_1 - {\bf x}_2\right)} &\delta_{ij},&\\
\langle {\gamma_{ij}(\bf{x_1})}{\gamma_{kl}(\bf{x_2})} \rangle &=& &\int \frac {d^3{\bf k}}{(2\pi)^3}{1\over {\bf k}^2} e^{i{\bf k}\cdot\left({\bf x}_1 - {\bf x}_2\right)} &P_{ij;kl}&,
\end{array}
\ee
where $P_{ij;kl}={1\over 2}\left(\delta_{ik}\delta_{jl}+\delta_{il}\delta_{jk}-2\delta_{ij}\delta_{kl}\right)$. Note that the propagator of the gravitomagnetic vector field, which is the most common in this interaction as we shall see, is diagonal in the harmonic gauge -- an advantageous feature of the NRG parametrization.

Before we specify the calculation, we need to consider the gauge for the redundant spin degrees of freedom. We may choose the  Newton-Wigner (NW) coordinates, which yield the spin supplementary condition (SSC) given in the local Lorentz frame by \cite{Hanson:1974qy}
\be \label{eq:hrnw}
mS^{0\mu}= S^{\mu\nu}p_\nu. 
\ee 
A covariant form of the NW SSC as appropriate for curved spacetime was introduced in \cite{Steinhoff:2008zr}; see Eq.~(4.9). We may also choose a conjugate gauge for the body fixed tetrads given by 
\be
e^\mu_0=\frac{p^\mu}{m}, 
\ee
so that in the rest frame we reduce to pure rotations as physically required \cite{Hanson:1974qy}. Using the constraints in Eq.~(\ref{eq:hrnw}) given in the local Lorentz frame we get 
\be\label{eq:s0} 
S^{i0} = \frac{1}{2}S^{ij}v^j + \frac{1}{4}S^{ij}A_{j} + O(v^4),  
\ee
where we express the subleading temporal spin entries $S^{i0}$, given in the local Lorentz frame, in terms of the coordinate velocity, and the metric field. Note the $O(v^3)$ subleading term appearing here, which is absent in flat spacetime, where the local Lorentz and coordinate velocities coincide, and only terms explicitly odd in the power of the velocity exist. Note also that it is the gravitomagnetic vector field, which is involved in this leading curved spacetime contribution to the temporal spin entries. 

We may also choose the covariant SSC, given by $S^{\mu\nu}p_\nu=0$ \cite{Hanson:1974qy}. However, to the order considered here this is equivalent to the condition given by $S^{\mu\nu}u_\nu=0$, i.e.~in the covariant SSC, we have for the spin temporal components
\be\label{eq:covst}
S^{i0} = S^{ij}v^j + \frac{1}{2}S^{ij}A_{j} + \cdots,
\ee
which again implies a suppression of the temporal spin components $S^{i0}$ by one order of $v$ with respect to the spatial components $S^{ij}$. However, here there is no infinite expansion explicit in $v$, yielding contributions to all orders, as opposed to Eq.~(\ref{eq:s0}). These constraints, e.g.~those expressed in Eqs.~(\ref{eq:s0}, \ref{eq:covst}), eliminate the redundant unphysical degrees of freedom, so that the spin may be represented by a 3-vector defined as $S^{ij} \equiv \epsilon^{ijk}S^{k}$.

\begin{figure}[t]
\includegraphics{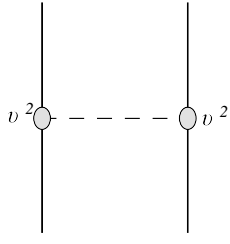}
\caption{Feynman diagram of the leading order spin1-spin2 interaction. The thick solid lines represent the time evolution of the point particles worldlines. The blobs represent spin insertions on the worldline. The dashed line represents the $A_i$ gravitomagnetic vector field propagator.}\label{lo}
\end{figure}

We go on to expand the spin part of the action in terms of the NRG fields to obtain the following leading order (LO) spin graviton coupling:  
\be 
L_2 = \frac{1}{4} S^{ij}F_{ij},
\ee
where the subscript is standard notation for the power of the orbital velocity of this term in the action. Thus, the LO spin1-spin2 potential follows from a one-graviton exchange of the gravitomagnetic vector graviton, with two of these LO spin vertices, and will scale accordingly as $v^4$ (2PN). The corresponding Feynman diagram is depicted in Fig.~1, resulting in the well-known LO spin1-spin2 potential given by
\be\label{eq:vlo}
V_{S_1S_2}^{LO} = -\frac{G}{r^3}(\vec S_1\cdot \vec S_2-3 \vec S_1\cdot \vec n\,\, \vec S_2\cdot \vec n),
\ee
where $r\equiv\left|{\vec{x}_1-\vec{x}_2}\right|$, and $\vec{n}\equiv\frac{\vec{r}}{r}$.

For the next order we must also take into account the subleading temporal spin entries (i.e.~$S^{i0}$), where we are considering the SSC on the level of the action ,which is accurate to the order we are calculating here \cite{Porto:2007px}. 
Considering vertices involving spin to order $v^3$, we have
\be\label{eq:l3}
L_3 = S^{ij}\partial_j\phi v^i + S^{0i}\partial_i\phi + \frac{1}{2}S^{ij}\partial_i\gamma_{jk}v^k, 
\ee
where the leading order term of the temporal spin entry $S^{0i}$ is considered in the second term here. Calculating to order $v^6$ (3PN), we should include diagrams with two insertions of $L_3$ as depicted in Figs.~2(a1), 2(a2). The values of these diagrams are given by
\bea 
Fig.~2(a1) = \frac{G}{r^3}&&\left[ \vec{S}_1\cdot\vec{S}_2\,\,\vec{v}_1\cdot\vec{v}_2-\vec{S}_1\cdot\vec{v}_2\,\,\vec{S}_2\cdot\vec{v}_1-3\left(\vec{S}_1\times\vec{v}_1\right)\cdot\vec{n}\,\,\left(\vec{S}_2\times\vec{v}_2\right)\cdot\vec{n} \right]\nonumber\\
+\frac{G}{r^3}&& \left[S_1^{0i}S_2^{0i} - 3S_1^{0i}n^iS_2^{0j}n^j \right.\nonumber\\
&& \left. + S_1^{0i}\left((\vec{S}_2\times\vec{v}_2)^i - 3n^i\,\vec{S}_2\times\vec{v}_2\cdot\vec{n}\right) + S_2^{0i}\left((\vec{S}_1\times\vec{v}_1)^i - 3n^i\,\vec{S}_1\times\vec{v}_1\cdot\vec{n}\right) \right],\\
Fig.~2(a2) = \frac{G}{r^3}&&\left[ -2\vec{S}_1\cdot\vec{S}_2\,\,\vec{v}_1\cdot\vec{v}_2-\vec{S}_1\cdot\vec{v}_1\,\,\vec{S}_2\cdot\vec{v}_2+2\vec{S}_1\cdot\vec{v}_2\,\,\vec{S}_2\cdot\vec{v}_1+3\vec{S}_1\cdot\vec{n}\,\,\vec{S}_2\cdot\vec{n}\,\,\vec{v}_1\cdot\vec{v}_2 \right.\nonumber\\
&&
\left.+6\left(\vec{S}_1\times\vec{v}_1\right)\cdot\vec{n}\,\,\left(\vec{S}_2\times\vec{v}_2\right)\cdot\vec{n}-3\left(\vec{S}_1\times\vec{v}_2\right)\cdot\vec{n}\,\,\left(\vec{S}_2\times\vec{v}_1\right)\cdot\vec{n} \right].
\eea
Here, a multiplicative factor of $\int dt$ is suppressed in the values of diagrams, and is to be further omitted from all diagram values. 

For the NLO we should also include time derivatives, thereby departing from stationarity. At order $v^4$ we must include 
\be\label{eq:l4}
L_4 = -\frac{1}{2}S^{0i}\partial_iA_jv^j + \frac{1}{2}S^{0i}\partial_0A_i. 
\ee
Both terms contain temporal spin entries $S^{0i}$ considered in their LO. Note the second term here, first containing time dependence of the potential field, that of the gravitomagnetic vector. Thus, we should also include diagrams with one insertion of $L_4$ and the LO spin vertex $L_2$ as depicted in Fig.~2(b), which equals 
\bea\label{eq:cd}
Fig.~2(b)= \frac{G}{r^3}&&\left[
3S_1^{0i}\left(n^i\vec{S}_2\times\vec{v}_1\cdot\vec{n}-(\vec{S}_2\times\vec{n})^i\vec{v}_1\cdot\vec{n}\right) + 3S_2^{0i}\left(n^i\vec{S}_1\times\vec{v}_2\cdot\vec{n}-(\vec{S}_1\times\vec{n})^i\vec{v}_2\cdot\vec{n}\right)\right]\nonumber\\
+ \frac{G}{r^2}&&\left[
\partial_tS_1^{0i}(\vec{S}_2\times\vec{n})^i\ - \partial_tS_2^{0i}(\vec{S}_1\times\vec{n})^i\right].
\eea
However, note that due to the time derivative term this diagram may be evaluated in two ways. The one we took here is straightforward, but yields acceleration terms. Such terms may seem undesirable since they call for the use of the equations of motion (EOM) at the level of the Lagrangian, a procedure which is known to be incorrect in many (in)famous examples. However, as originally noted by \cite{Schafer:1984}, and later formally evolved and treated, e.g.~in \cite{Damour:1990jh}, a substitution of low order EOM in higher order terms in the level of the Lagrangian is a correct procedure in GR, and is equivalent to performing a coordinate transformation. For example, if we consider the NW SSC, i.e.~Eq.~(\ref{eq:s0}) at the level of the action, we evaluate Fig.~2(b) according to Eq.~(\ref{eq:cd}) as
\bea\label{eq:cd1}
Fig.~2(b)= \frac{G}{2r^3}&&\left[
-3\vec{S}_1\cdot\vec{S}_2\left((\vec{v}_1\cdot\vec{n})^2+(\vec{v}_2\cdot\vec{n})^2\right)
+3\vec{S}_1\cdot\vec{v}_2\vec{S}_2\cdot\vec{n}\vec{v}_2\cdot\vec{n}+3\vec{S}_2\cdot\vec{v}_1\vec{S}_1\cdot\vec{n}\vec{v}_1\cdot\vec{n}\right.\nonumber\\
&&
\left.+3(\vec{S}_1\times\vec{v}_1)\cdot\vec{n}(\vec{S}_2\times\vec{v}_1)\cdot\vec{n}+3(\vec{S}_1\times\vec{v}_2)\cdot\vec{n}(\vec{S}_2\times\vec{v}_2)\cdot\vec{n}\right]\nonumber\\
&&
+\frac{G}{2r^2}\left[(\vec{S}_1\times\vec{a}_1)\cdot(\vec{S}_2\times\vec{n})-(\vec{S}_2\times\vec{a}_2)\cdot(\vec{S}_1\times\vec{n})\right]\nonumber\\
= \frac{G}{2r^3}&&\left[
-3\vec{S}_1\cdot\vec{S}_2\left((\vec{v}_1\cdot\vec{n})^2+(\vec{v}_2\cdot\vec{n})^2\right)
+3\vec{S}_1\cdot\vec{v}_2\vec{S}_2\cdot\vec{n}\vec{v}_2\cdot\vec{n}+3\vec{S}_2\cdot\vec{v}_1\vec{S}_1\cdot\vec{n}\vec{v}_1\cdot\vec{n}\right.\nonumber\\
&&
\left.+3(\vec{S}_1\times\vec{v}_1)\cdot\vec{n}(\vec{S}_2\times\vec{v}_1)\cdot\vec{n}+3(\vec{S}_1\times\vec{v}_2)\cdot\vec{n}(\vec{S}_2\times\vec{v}_2)\cdot\vec{n}\right]\nonumber\\
&&
-\frac{G^2(m_1+m_2)}{2r^4}\left[\vec{S}_1\cdot\vec{S}_2-\vec{S}_1\cdot\vec{n}\,\,\vec{S}_2\cdot\vec{n}\right].  
\eea
Here, we eliminated the acceleration terms using the LO EOM, given by  
\be
\vec{a}\equiv\vec{a}_1-\vec{a}_2=-\frac{G(m_1+m_2)}{r^2}\vec{n},  
\ee
resulting in the nonlinear term in $G$ appearing in the last line of Eq.~(\ref{eq:cd1}). By this substitution we implicitly have coordinate transformations made (leading us to the result in ADM coordinates). 
\begin{figure}[b]
\includegraphics{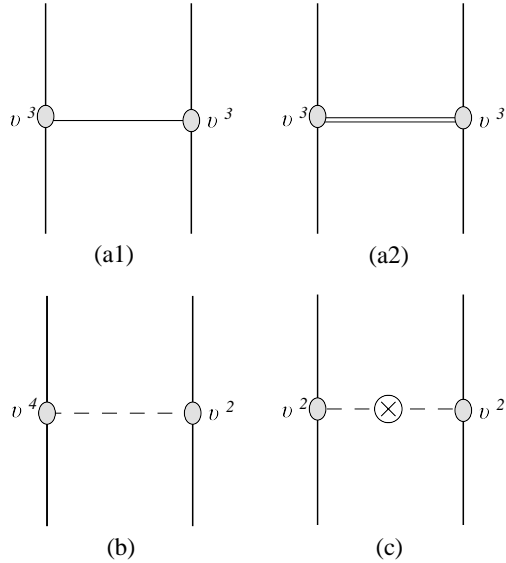}
\caption{Next-to-leading order spin1-spin2 interaction Feynman diagrams of one-graviton exchange. The solid line represents the scalar field propagator. The double line represents the 2-tensor field propagator. The cross vertex corresponds to an insertion of the graviton kinetic term. (b) should be included together with its mirror image.}\label{lnlo}
\end{figure}

Alternatively, the time derivative term from Eq.~(\ref{eq:l4}) may be evaluated by flipping the time derivative between the two PP worldlines, namely, by using the identity
\be
\int dt_1dt_2~\partial_{t_1}\delta(t_1-t_2)f(t_1)g(t_2)=-\int dt_1dt_2~\partial_{t_2}\delta(t_1-t_2)f(t_1)g(t_2).
\ee
Such an evaluation of the diagram in Fig.~2(b) avoids the acceleration terms, and the use of the EOM. Again, considering the NW SSC at the level of the action, it yields the following seemingly different value for the diagram:
\bea
Fig.~2(b)_{alt}= \frac{G}{2r^3}&&\left[ 2\vec{S}_1\cdot\vec{S}_2\,\,\vec{v}_1\cdot\vec{v}_2 -2\vec{S}_1\cdot\vec{v}_2\,\,\vec{S}_2\cdot\vec{v}_1 +\vec{S}_1\cdot\vec{v}_1\,\,\vec{S}_2\cdot\vec{v}_1
+\vec{S}_1\cdot\vec{v}_2\,\,\vec{S}_2\cdot\vec{v}_2\right.\nonumber\\
&&
-\vec{S}_1\cdot\vec{S}_2\left(\vec{v}_1^2+\vec{v}_2^2\right)-6\vec{S}_1\cdot\vec{S}_2\vec{v}_1\cdot\vec{n}\vec{v}_2\cdot\vec{n} +3\vec{S}_1\cdot\vec{v}_2\vec{S}_2\cdot\vec{n}\vec{v}_1\cdot\vec{n}+3\vec{S}_2\cdot\vec{v}_1\vec{S}_1\cdot\vec{n}\vec{v}_2\cdot\vec{n}\nonumber\\
&&
\left.+3(\vec{S}_1\times\vec{v}_1)\cdot\vec{n}(\vec{S}_2\times\vec{v}_1)\cdot\vec{n}+3(\vec{S}_1\times\vec{v}_2)\cdot\vec{n}(\vec{S}_2\times\vec{v}_2)\cdot\vec{n}\right]. 
\eea
This was the route taken in \cite{Porto:2006bt}, thus requiring an explicit canonical transformation to reproduce the equivalent result in the ADM coordinates after an addition of a missing piece in the potential as shown later in  \cite{Porto:2007tt}. The time derivative term of Eq.~(\ref{eq:l4}) also yields spin precession terms, thus similarly requiring the use of the LO spin EOM, which scales as \cite{Porto:2005ac} 
\be 
\frac{dS}{dt}\sim \frac{Gm}{r^2} Sv.
\ee
This yields a 4PN contribution, and can therefore be taken here as zero.

There is a further contribution from the one-graviton exchange sector, which arises from the leading correction to the potential graviton propagators, i.e.~the graviton kinetic terms in the gravitational action. It is a first departure from the stationarity approximation in the gravitational action. Extracting from the EH action the quadratic time dependent terms, which are relevant to the order we are calculating here, we have 
\be\label{eq:2ehc}
S^{g^2}_{EH} \supset -\frac{1}{16\pi G}\int dt d^3x\left[ 4\partial_iA_i\partial_0\phi +\partial_0A_i\partial_j \gamma_{ij}-\frac{1}{2}\partial_0A_i\partial_i \gamma_{jj}-\frac{1}{2}\partial_iA_i\partial_0 \gamma_{jj} \right].
\ee
Thus, in order to reduce the number of contributing diagrams as much as possible, we refine the gauge fixing term of Eq.~(\ref{eq:gf}) to eliminate undesired quadratic vertices. Hence, we reset it to be 
\be
S_{\it{GF}} = \frac{1}{32\pi G} \int dtd^3x\left[\left( \partial_{i}A_i+\left(4\partial_0\phi -\frac{1}{2}\partial_0\gamma_{ii}\right)\right)^2 - \left(\left(\partial_j\gamma_{ij}-\frac{1}{2}\partial_i\gamma_{jj}\right)-\partial_0A_i\right)^2\right], 
\ee
so that to the order considered here, we are left with the following time dependent quadratic term: 
\be\label{eq:2gc} 
\left(S_{EH}+S_{GF}\right)\supset -\frac{1}{32\pi G}\int d^4x (\partial_0A_i)^2.
\ee
Hence, the leading deviation from stationarity in the gravitational self-interaction in the spin1-spin2 sector involves again the gravitomagnetic vector field alone. However, note that since the gravitomagnetic vector has no kinetic term in the original EH action, this contribution arises as a pure gauge fixing correction, i.e.~it does not involve corrections to the KK reduced action. This last contribution is depicted in the diagram shown in Fig.~2(c), which equals
\bea 
Fig.~2(c)=\frac{G}{2r^3}&&\left[  -\vec{S}_1\cdot\vec{S}_2\,\,\vec{v}_1\cdot\vec{v}_2 +\vec{S}_1\cdot\vec{v}_1\,\,\vec{S}_2\cdot\vec{v}_2
+\vec{S}_1\cdot\vec{v}_2\,\,\vec{S}_2\cdot\vec{v}_1
+3\vec{S}_1\cdot\vec{S}_2\vec{v}_1\cdot\vec{n}\vec{v}_2\cdot\vec{n}
-3\vec{S}_1\cdot\vec{n}\vec{S}_2\cdot\vec{n}\vec{v}_1\cdot\vec{v}_2\right.\nonumber\\
&&
-3\vec{S}_1\cdot\vec{v}_1\vec{S}_2\cdot\vec{n}\vec{v}_2\cdot\vec{n}
-3\vec{S}_2\cdot\vec{v}_2\vec{S}_1\cdot\vec{n}\vec{v}_1\cdot\vec{n}
-3\vec{S}_1\cdot\vec{v}_2\vec{S}_2\cdot\vec{n}\vec{v}_1\cdot\vec{n}
-3\vec{S}_2\cdot\vec{v}_1\vec{S}_1\cdot\vec{n}\vec{v}_2\cdot\vec{n}\nonumber\\
&&
+\left.15\vec{S}_1\cdot\vec{n}\vec{S}_2\cdot\vec{n}\vec{v}_1\cdot\vec{n}\vec{v}_2\cdot\vec{n}\right].
\eea
Here, one also makes use of the LO spin EOM to eliminate precession terms.

At order 3PN nonlinear contributions in $G$ must also be included. First, we have the two-graviton exchange (including only connected diagrams after stripping off worldlines). The relevant terms in the Lagrangian, which are quadratic in the metric, are given by   
\be\label{eq:lq} 
L_4^{g^2} = S^{ij}F_{ij}\phi - \frac{1}{2}S^{ij}A_i\partial_j\phi. 
\ee 
However, it is important to note that this sector contains a further contribution from the subleading $O(v^3)$ term of the $S^{0i}$ spin entries, e.g.~in Eqs.~(\ref{eq:s0}, \ref{eq:covst}), which incorporates the effect of curved spacetime. It arises as the NLO substitution in the $S^{0i}$ term in Eq.~(\ref{eq:l3}). For example, if we consider the NW SSC, it equals
\be\label{eq:lqssc}
L_{4(SSC)}^{g^2} = \frac{1}{4}S^{ij}A_i\partial_j\phi. 
\ee
This contribution was considered only in \cite{Porto:2007tt}, and then regarded as an additional spin-orbit effect, but here it is incorporated into the Feynman diagrams as being a proper spin1-spin2 contribution. Hereby, we consistently consider the SSC at the level of the action, a procedure accurate to the order calculated here as we noted already, while not being generally valid. 
\begin{figure}[t]
\includegraphics{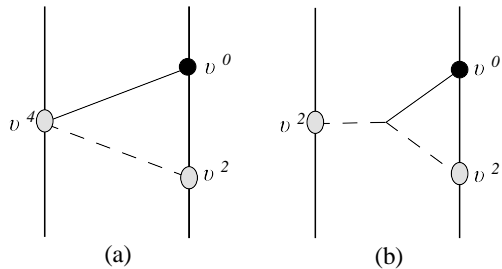}
\caption{Feynman diagrams of nonlinear spin1-spin2 interaction in next-to-leading order. The black blobs represent mass vertices on the worldline. These diagrams should be included together with their mirror images.}\label{nlnlo}
\end{figure}
The terms in Eqs.~(\ref{eq:lq}, \ref{eq:lqssc}) scale as $v^4$, so that diagrams containing them must also include a LO spin insertion, and a LO mass insertion from the nonspinning part of the PP action given by 
\be 
L_0 = -m \phi. 
\ee
This diagram is shown in Fig.~3(a), and considering Eq.~(\ref{eq:lq}), it equals 
\be
Fig.~3(a)= -\frac{G^2(m_1+m_2)}{r^4}\left[5\vec{S}_1\cdot\vec{S}_2-13\vec{S}_1\cdot\vec{n}\,\,\vec{S}_2\cdot\vec{n}\right].
\ee
In addition, the contribution from e.g.~Eq.~(\ref{eq:lqssc}), arising from the NW SSC, yields
\be \label{eq:lqsscadd}
Fig.~3(a)_{SSC}= \frac{G^2(m_1+m_2)}{2r^4}\left[\vec{S}_1\cdot\vec{S}_2-\vec{S}_1\cdot\vec{n}\,\,\vec{S}_2\cdot\vec{n}\right].
\ee

Finally, we must include the contribution from those diagrams including the three-graviton vertex, which scales as $v^2$ \cite{Goldberger:2004jt}. To the order considered here diagrams including this vertex should therefore include two LO spin insertions and one LO mass insertion, with the three-graviton vertex given by the following cubic term from the KK reduced action: 
\be
S_{KK}\supset \frac{1}{8\pi G}\int dt d^3x~\left[\phi \partial_iA_j\left(\partial_iA_j-\partial_jA_i\right)\right].
\ee
This vertex is easily read from the KK reduced action, already making computation much faster compared to the tedious extraction of the cubic part of the EH action, which contains a very large number of terms, each with a complicated tensor index structure as in \cite{Goldberger:2004jt}. The corresponding diagram is depicted in Fig.~3(b). This diagram contains a loop integral, and requires regularization. It is handled with dimensional regularization by the usual techniques, see e.g.~\cite{Collins:1984xc}. The diagram is evaluated as
\be
Fig.~3(b) = \frac{2G^2(m_1+m_2)}{r^4}\left[\vec{S}_1\cdot\vec{S}_2-2\vec{S}_1\cdot\vec{n}\,\,\vec{S}_2\cdot\vec{n}\right].
\ee
The long calculation required in \cite{Porto:2006bt}, calling for the aid of a symbolic manipulation software, becomes extremely feasible by hand as noted already in \cite{Kol:2007bc}. Both diagrams of the nonlinear interaction are included with their mirror images.

\section{Next-to-leading order spin1-spin2 potential}

Summing up all of the contributions from the diagrams, we obtain the NLO spin1-spin2 Lagrangian. Here, we give the result without substituting in the SSC, leaving the $S^{0i}$ entries as independent degrees of freedom, and without taking into account the SSC field contribution, e.g.~in Eq.~(\ref{eq:lqssc}). Next, we will consider both the implementation of the NW SSC at the action level, and the covariant SSC at the Hamiltonian level. The results are to be compared with Eq.~(2.11) of \cite{Steinhoff:2007mb}, and we will show that both ways lead to an agreement with the result obtained in the ADM Hamiltonian formalism. Our Lagrangian then reads
\bea\label{eq:ssb}
L_{S_1S_2}^{NLO} = -\frac{G}{2r^3}&&\left[ \vec{S}_1\cdot
\vec{S}_2\left(3\vec{v}_1\cdot\vec{v}_2-3\vec{v}_1\cdot{\vec{n}}\vec{v}_2\cdot{\vec{n}}\right)
+\vec{S}_1\cdot\vec{v}_1\vec{S}_2\cdot\vec{v}_2-3\vec{S}_1\cdot\vec{v}_2\vec{S}_2\cdot\vec{v}_1
-3\vec{S}_1\cdot\vec{n}\vec{S}_2\cdot\vec{n}\left(\vec{v}_1\cdot\vec{v}_2+5\vec{v}_1\cdot\vec{n}\vec{v}_2\cdot\vec{n}\right)\right.\nonumber\\
&&
+3\vec{S}_1\cdot\vec{v}_1\vec{S}_2\cdot\vec{n}\vec{v}_2\cdot\vec{n}
+3\vec{S}_2\cdot\vec{v}_2\vec{S}_1\cdot\vec{n}\vec{v}_1\cdot\vec{n}
+3\vec{S}_1\cdot\vec{v}_2\vec{S}_2\cdot\vec{n}\vec{v}_1\cdot\vec{n}
+3\vec{S}_2\cdot\vec{v}_1\vec{S}_1\cdot\vec{n}\vec{v}_2\cdot\vec{n}\nonumber\\
&&\left. 
-6(\vec{S}_1\times\vec{v}_1)\cdot\vec{n}(\vec{S}_2\times\vec{v}_2)\cdot\vec{n}
+6(\vec{S}_1\times\vec{v}_2)\cdot\vec{n}(\vec{S}_2\times\vec{v}_1)\cdot\vec{n}\right]
-3\frac{G^2(m_1+m_2)}{r^4}\left[\vec{S}_1\cdot\vec{S}_2-3\vec{S}_1\cdot\vec{n}\vec{S}_2\cdot\vec{n}\right]\nonumber\\
+\frac{G}{r^3}&&\left[S_1^{0i}S_2^{0i} - 3S_1^{0i}n^iS_2^{0j}n^j + S_1^{0i}\left((\vec{S}_2\times\vec{v}_2)^i - 3n^i\vec{S}_2\times\vec{v}_2\cdot\vec{n}\right) + S_2^{0i}\left((\vec{S}_1\times\vec{v}_1)^i - 3n^i\vec{S}_1\times\vec{v}_1\cdot\vec{n}\right) \right.\nonumber\\
&&+\left.
3S_1^{0i}\left(n^i\vec{S}_2\times\vec{v}_1\cdot\vec{n}-(\vec{S}_2\times\vec{n})^i\vec{v}_1\cdot\vec{n}\right) + 3S_2^{0i}\left(n^i\vec{S}_1\times\vec{v}_2\cdot\vec{n}-(\vec{S}_1\times\vec{n})^i\vec{v}_2\cdot\vec{n}\right)\right]\nonumber\\
+ &&\frac{G}{r^2}\left[
\partial_tS_1^{0i}(\vec{S}_2\times\vec{n})^i - \partial_tS_2^{0i}(\vec{S}_1\times\vec{n})^i\right],
\eea
where the SSC dependent part appears separately in the last three lines of the above expression.

\subsection{Implementation of NW SSC at the action level} \label{subsec:nw}

If we consider the NW SSC at the level of the action, i.e.~we substitute in the $S^{i0}$ entries from Eq.~(\ref{eq:s0}), and take into account the contribution from Eqs.~(\ref{eq:lqssc}, \ref{eq:lqsscadd}), we obtain
\bea\label{eq:ss}
L_{S_1S_2}^{NLO} = -\frac{G}{2r^3}&&\left[ \vec{S}_1\cdot
\vec{S}_2\left(\frac{1}{2}\vec{v}_1\cdot\vec{v}_2-3\vec{v}_1\cdot{\vec{n}}\vec{v}_2\cdot{\vec{n}}
+3\left((\vec{v}_1\cdot{\vec{n}})^2+(\vec{v}_2\cdot{\vec{n}})^2\right)\right)
+\vec{S}_1\cdot\vec{v}_1\vec{S}_2\cdot\vec{v}_2-\frac{1}{2}\vec{S}_1\cdot\vec{v}_2\vec{S}_2\cdot\vec{v}_1\right.\nonumber\\
&&
-3\vec{S}_1\cdot\vec{n}\vec{S}_2\cdot\vec{n}
\left(\vec{v}_1\cdot\vec{v}_2+5\vec{v}_1\cdot\vec{n}\vec{v}_2\cdot\vec{n}\right)
+3\vec{S}_1\cdot\vec{v}_1\vec{S}_2\cdot\vec{n}\vec{v}_2\cdot\vec{n}
+3\vec{S}_2\cdot\vec{v}_2\vec{S}_1\cdot\vec{n}\vec{v}_1\cdot\vec{n}
\nonumber\\
&& 
+3\vec{S}_1\cdot\vec{v}_2\vec{S}_2\cdot\vec{n}\vec{v}_1\cdot\vec{n}
+3\vec{S}_2\cdot\vec{v}_1\vec{S}_1\cdot\vec{n}\vec{v}_2\cdot\vec{n}
-3\vec{S}_1\cdot\vec{v}_2\vec{S}_2\cdot\vec{n}\vec{v}_2\cdot\vec{n}
-3\vec{S}_2\cdot\vec{v}_1\vec{S}_1\cdot\vec{n}\vec{v}_1\cdot\vec{n}
\nonumber\\
&&
-3(\vec{S}_1\times\vec{v}_1)\cdot\vec{n}(\vec{S}_2\times\vec{v}_1)\cdot\vec{n}
-3(\vec{S}_1\times\vec{v}_2)\cdot\vec{n}(\vec{S}_2\times\vec{v}_2)\cdot\vec{n}
+\frac{3}{2}(\vec{S}_1\times\vec{v}_1)\cdot\vec{n}(\vec{S}_2\times\vec{v}_2)\cdot\vec{n}\nonumber\\
&&\left.
+6(\vec{S}_1\times\vec{v}_2)\cdot\vec{n}(\vec{S}_2\times\vec{v}_1)\cdot\vec{n}
\right] 
+\frac{G}{2r^2}\left[(\vec{S}_1\times\vec{a}_1)\cdot(\vec{S}_2\times\vec{n})-(\vec{S}_2\times\vec{a}_2)\cdot(\vec{S}_1\times\vec{n})\right]\nonumber\\
&&
-\frac{G^2 (m_1+m_2)}{2r^4}\left[5\vec{S}_1\cdot\vec{S}_2-17\vec{S}_1\cdot\vec{n}\vec{S}_2\cdot\vec{n}\right]\nonumber\\
=-\frac{G}{2r^3}&&\left[\cdots\right]-3\frac{G^2 (m_1+m_2)}{r^4}\left[\vec{S}_1\cdot\vec{S}_2-3\vec{S}_1\cdot\vec{n}\vec{S}_2\cdot\vec{n}\right],
\eea
where in the last equality we have used the LO EOM for the accelerations. 

We should make a Legendre transformation with respect to the coordinate velocities in order to obtain the NLO spin1-spin2 Hamiltonian. The relevant part of the Lagrangian reads
\be\label{eq:lreq}  
L = L_N-V_{SO}^{LO}-V_{S_1S_2}^{LO}+L_{S_1S_2}^{NLO},  
\ee
where $L_N$ is the Newtonian Lagrangian given by 
\be
L_N = \frac{1}{2}\sum_{a=1}^{2}m_a\vec{v}_a^2+\frac{Gm_1m_2}{r}.
\ee
$V_{SO}^{LO}$ is the LO spin-orbit potential, given by (e.g.~from \cite{Steinhoff:2007mb})
\be\label{eq:vsonw}
V_{SO}^{LO} = -\frac{Gm_2}{r^2} \left[{\vec{S}}_1\cdot\left({\vec{v}}_1\times{\vec{n}} - 2{\vec{v}}_2\times{\vec{n}}\right) + S_1^{0i}n^i\right] + [1 \leftrightarrow 2]
= -\frac{Gm_2}{r^2} \vec{S}_1\cdot \left(\frac{3}{2}\vec{v}_1\times\vec{n}-2\vec{v}_2\times\vec{n}\right) + [1 \leftrightarrow 2],
\ee
where in the second equality the NW SSC from Eq.~(\ref{eq:s0}) is implemented. $V_{S_1S_2}^{LO}$ is the LO spin1-spin2 coupling given here in Eq.~(\ref{eq:vlo}), and $L_{S_1S_2}^{NLO}$ is the NLO spin1-spin2 coupling that we calculate here. Note that the LO spin-orbit potential already requires the use of SSC, so we need to pay special attention to it. The canonical momenta are given by 
\be 
\vec{p}_a = \frac{\partial L}{\partial \vec{v}_a},
\ee
so that e.g.
\be\label{eq:pc} 
\vec{p}_1 = m_1\vec{v}_1 - 
\frac{G}{r^2}\left(\frac{3}{2}m_2(\vec{S}_1\times\vec{n})+2m_1(\vec{S}_2\times\vec{n})\right) + 
\frac{\partial L_{S_1S_2}^{NLO}}{\partial \vec{v}_1}.
\ee
Consequently, we have for the velocity, e.g. 
\be\label{eq:vc}
\vec{v}_1 = \frac{\vec{p}_1}{m_1} + \frac{G}{m_1r^2}\left(\frac{3}{2}m_2(\vec{S}_1\times\vec{n})+2m_1(\vec{S}_2\times\vec{n})\right)-\frac{
\frac{\partial L_{S_1S_2}^{NLO}}{\partial \vec{v}_1}}{m_1}. 
\ee
The Hamiltonian is obtained by substituting in the velocities given in terms of canonical momenta in the following expression:
\be
H = \sum_{a=1,2}\vec{v}_a\cdot\vec{p}_a - L.
\ee 
Note that the $\frac{\partial L_{S_1S_2}^{NLO}}{\partial \vec{v}_a}$ terms get canceled, and do not contribute here.

The Hamiltonian then takes the following form: 
\be
H = H_O+H_{SO}+H_{SS},
\ee 
where $H_O$ is a part of the orbital Hamiltonian, $H_{SO}$ is a part of the spin-orbit (linear in spins) Hamiltonian, and $H_{SS}$ is the spin-spin (quadratic in spins) sector of the Hamiltonian up to NLO. Thus, the NLO spin1-spin2 Hamiltonian equals
\be
H_{S_1S_2}^{NLO} = -L_{S_1S_2}^{NLO} + \frac{3G^2 (m_1+m_2)}{r^4}\left[\vec{S}_1\cdot\vec{S}_2-\vec{S}_1\cdot\vec{n}\vec{S}_2\cdot\vec{n}\right],
\ee
and finally we obtain
\bea\label{eq:hss}
H_{S_1S_2}^{NLO} = \frac{G}{2r^3}&&\left[\frac{1}{m_1m_2}\left(\frac{1}{2}\vec{S}_1\cdot\vec{S}_2\vec{p}_1\cdot\vec{p}_2
-3\vec{S}_1\cdot\vec{S}_2\vec{p}_1\cdot{\vec{n}}\vec{p}_2\cdot{\vec{n}}
+\vec{S}_1\cdot\vec{p}_1\vec{S}_2\cdot\vec{p}_2
-\frac{1}{2}\vec{S}_1\cdot\vec{p}_2\vec{S}_2\cdot\vec{p}_1\right.\right.\nonumber\\
&&
-3\vec{S}_1\cdot\vec{n}\vec{S}_2\cdot\vec{n}\vec{p}_1\cdot\vec{p}_2
-15\vec{S}_1\cdot\vec{n}\vec{S}_2\cdot\vec{n}\vec{p}_1\cdot\vec{n}\vec{p}_2\cdot\vec{n}
+3\vec{S}_1\cdot\vec{p}_1\vec{S}_2\cdot\vec{n}\vec{p}_2\cdot\vec{n}
+3\vec{S}_2\cdot\vec{p}_2\vec{S}_1\cdot\vec{n}\vec{p}_1\cdot\vec{n}
\nonumber\\
&&\left.
+3\vec{S}_1\cdot\vec{p}_2\vec{S}_2\cdot\vec{n}\vec{p}_1\cdot\vec{n}
+3\vec{S}_2\cdot\vec{p}_1\vec{S}_1\cdot\vec{n}\vec{p}_2\cdot\vec{n}
+\frac{3}{2}(\vec{S}_1\times\vec{p}_1)\cdot\vec{n}(\vec{S}_2\times\vec{p}_2)\cdot\vec{n}
+6(\vec{S}_1\times\vec{p}_2)\cdot\vec{n}(\vec{S}_2\times\vec{p}_1)\cdot\vec{n}\right)\nonumber\\
&&+\frac{3}{m_1^2}\left(
\vec{S}_1\cdot\vec{S}_2(\vec{p}_1\cdot{\vec{n}})^2
-\vec{S}_2\cdot\vec{p}_1\vec{S}_1\cdot\vec{n}\vec{p}_1\cdot\vec{n}
-(\vec{S}_1\times\vec{p}_1)\cdot\vec{n}(\vec{S}_2\times\vec{p}_1)\cdot\vec{n}\right)\nonumber\\
&&\left.+\frac{3}{m_2^2}\left(
\vec{S}_1\cdot\vec{S}_2(\vec{p}_2\cdot{\vec{n}})^2
-\vec{S}_1\cdot\vec{p}_2\vec{S}_2\cdot\vec{n}\vec{p}_2\cdot\vec{n}
-(\vec{S}_1\times\vec{p}_2)\cdot\vec{n}(\vec{S}_2\times\vec{p}_2)\cdot\vec{n}\right)\right]\nonumber\\
&&+\frac{G^2 (m_1+m_2)}{r^4}\left[6\vec{S}_1\cdot\vec{S}_2-12\vec{S}_1\cdot\vec{n}\vec{S}_2\cdot\vec{n}\right].
\eea
This is just equal to the result given in Eq.~(2.11) of \cite{Steinhoff:2007mb}, obtained by the ADM Hamiltonian formalism.  

\subsection{Implementation of the covariant SSC at the Hamiltonian level}

Let us consider our NLO spin1-spin2 interaction Lagrangian given in Eq.~(\ref{eq:ssb}) with the implementation of the covariant SSC only at the Hamiltonian level, i.e. considering the $S^{i0}$ as independent degrees of freedom until the obtainment of a Hamiltonian, and only then implementing the covariant SSC. 

Again, we need to Legendre transform Eq.~(\ref{eq:ssb}), but now instead of Eqs.~(\ref{eq:pc}, \ref{eq:vc}), we will have from Eq.~(\ref{eq:vsonw}) e.g.
\be\label{eq:pcc} 
\vec{p}_1 = m_1\vec{v}_1 - 
\frac{G}{r^2}\left(m_2(\vec{S}_1\times\vec{n})+2m_1(\vec{S}_2\times\vec{n})\right) + 
\frac{\partial L_{S_1S_2}^{NLO}}{\partial \vec{v}_1},
\ee
\be\label{eq:vcc}
\vec{v}_1 = \frac{\vec{p}_1}{m_1} + \frac{G}{m_1r^2}\left(m_2(\vec{S}_1\times\vec{n})+2m_1(\vec{S}_2\times\vec{n})\right)-\frac{
\frac{\partial L_{S_1S_2}^{NLO}}{\partial \vec{v}_1}}{m_1}. 
\ee

Thus, Eq.~(\ref{eq:ssb}) is Legendre transformed into the following NLO spin1-spin2 Hamiltonian: 
\bea\label{eq:hs1s2}
H_{S_1S_2}^{NLO}&&= -L_{S_1S_2}^{NLO} + \frac{2G^2 (m_1+m_2)}{r^4}\left[\vec{S}_1\cdot\vec{S}_2-\vec{S}_1\cdot\vec{n}\vec{S}_2\cdot\vec{n}\right] 
+\frac{G}{r^2}\left[m_2 S_1^{i0}n^i + (1 \leftrightarrow 2)\right]\nonumber\\
&&= \frac{G}{2r^3}\left[\frac{1}{m_1m_2}\left(\vec{S}_1\cdot
\vec{S}_2\left(3\vec{p}_1\cdot\vec{p}_2-3\vec{p}_1\cdot{\vec{n}}\vec{p}_2\cdot{\vec{n}}\right)
+\vec{S}_1\cdot\vec{p}_1\vec{S}_2\cdot\vec{p}_2-3\vec{S}_1\cdot\vec{p}_2\vec{S}_2\cdot\vec{p}_1\right.\right.\nonumber\\
&&-3\vec{S}_1\cdot\vec{n}\vec{S}_2\cdot\vec{n}
\left(\vec{p}_1\cdot\vec{p}_2+5\vec{p}_1\cdot\vec{n}\vec{p}_2\cdot\vec{n}\right)
+3\vec{S}_1\cdot\vec{p}_1\vec{S}_2\cdot\vec{n}\vec{p}_2\cdot\vec{n}
+3\vec{S}_2\cdot\vec{p}_2\vec{S}_1\cdot\vec{n}\vec{p}_1\cdot\vec{n}
\nonumber\\
&&\left.\left. +3\vec{S}_1\cdot\vec{p}_2\vec{S}_2\cdot\vec{n}\vec{p}_1\cdot\vec{n}
+3\vec{S}_2\cdot\vec{p}_1\vec{S}_1\cdot\vec{n}\vec{p}_2\cdot\vec{n}
-6(\vec{S}_1\times\vec{p}_1)\cdot\vec{n}(\vec{S}_2\times\vec{p}_2)\cdot\vec{n}
+6(\vec{S}_1\times\vec{p}_2)\cdot\vec{n}(\vec{S}_2\times\vec{p}_1)\cdot\vec{n}\right)\right]\nonumber\\
&&-\frac{G}{r^3}\left[S_1^{0i}S_2^{0i} - 3S_1^{0i}n^iS_2^{0j}n^j + \frac{S_1^{0i}}{m_2}\left((\vec{S}_2\times\vec{p}_2)^i - 3n^i\vec{S}_2\times\vec{p}_2\cdot\vec{n}\right)+\frac{S_2^{0i}}{m_1}\left((\vec{S}_1\times\vec{p}_1)^i - 3n^i\vec{S}_1\times\vec{p}_1\cdot\vec{n}\right) \right.\nonumber\\
&&\left.+\frac{3S_1^{0i}}{m_1}\left(n^i\vec{S}_2\times\vec{p}_1\cdot\vec{n}-(\vec{S}_2\times\vec{n})^i\vec{p}_1\cdot\vec{n}\right)  +\frac{3S_2^{0i}}{m_2}\left(n^i\vec{S}_1\times\vec{p}_2\cdot\vec{n}-(\vec{S}_1\times\vec{n})^i\vec{p}_2\cdot\vec{n}\right)\right]\nonumber\\
&&+\frac{G^2(m_1+m_2)}{r^4}\left[5\vec{S}_1\cdot\vec{S}_2-11\vec{S}_1\cdot\vec{n}\vec{S}_2\cdot\vec{n}\right]\nonumber\\
&&-\frac{G}{r^2}\left[
\partial_tS_1^{0i}(\vec{S}_2\times\vec{n})^i - \partial_tS_2^{0i}(\vec{S}_1\times\vec{n})^i\right]
+\frac{G}{r^2}\left[m_2 S_1^{i0}n^i + (1 \leftrightarrow 2)\right].
\eea
Note the last term here originating from the SSC dependent part of the LO spin-orbit interaction in Eq.~(\ref{eq:vsonw}). We need to keep track of this term, as it contributes to the NLO spin1-spin2 interaction too. Next, we go on to insert the covariant SSC to our Hamiltonian, using Eqs.~(\ref{eq:covst}), (\ref{eq:vcc}), and the metric for spinning binary black holes in harmonic coordinates, e.g.~as in \cite{Steinhoff:2009hb}. It is well known that the variables then obtained will be noncanonical. However, we shall proceed to make noncanonical transformations that relate them to the canonical variables \cite{Hanson:1974qy}.

Hence, upon the insertion of the covariant SSC our noncanonical Hamiltonian reads
\bea\label{eq:hs1s2cov}
H_{S_1S_2}^{NLO} &&= \frac{G}{2r^3}\left[\frac{1}{m_1m_2} \left(-\vec{S}_1\cdot\vec{S}_2\left(3\vec{p}_1\cdot\vec{p}_2+3\vec{p}_1\cdot{\vec{n}}\vec{p}_2\cdot{\vec{n}}\right)
+\vec{S}_1\cdot\vec{p}_1\vec{S}_2\cdot\vec{p}_2+3\vec{S}_1\cdot\vec{p}_2\vec{S}_2\cdot\vec{p}_1\right.\right.\nonumber\\
&&
-3\vec{S}_1\cdot\vec{n}\vec{S}_2\cdot\vec{n}
\left(\vec{p}_1\cdot\vec{p}_2+5\vec{p}_1\cdot\vec{n}\vec{p}_2\cdot\vec{n}\right)\nonumber\\
&&
+3\vec{S}_1\cdot\vec{p}_1\vec{S}_2\cdot\vec{n}\vec{p}_2\cdot\vec{n}
+3\vec{S}_2\cdot\vec{p}_2\vec{S}_1\cdot\vec{n}\vec{p}_1\cdot\vec{n}
+3\vec{S}_1\cdot\vec{p}_2\vec{S}_2\cdot\vec{n}\vec{p}_1\cdot\vec{n}
+3\vec{S}_2\cdot\vec{p}_1\vec{S}_1\cdot\vec{n}\vec{p}_2\cdot\vec{n}\nonumber\\
&&\left.
+12(\vec{S}_1\times\vec{p}_1)\cdot\vec{n}(\vec{S}_2\times\vec{p}_2)\cdot\vec{n}
+6(\vec{S}_1\times\vec{p}_2)\cdot\vec{n}(\vec{S}_2\times\vec{p}_1)\cdot\vec{n}
\right)\nonumber\\
&& +\frac{6}{m_1^2}\left(
\vec{S}_1\cdot\vec{S}_2(\vec{p}_1\cdot{\vec{n}})^2
-\vec{S}_2\cdot\vec{p}_1\vec{S}_1\cdot\vec{n}\vec{p}_1\cdot\vec{n}
-(\vec{S}_1\times\vec{p}_1)\cdot\vec{n}(\vec{S}_2\times\vec{p}_1)\cdot\vec{n}
\right)\nonumber\\
&& \left.+\frac{6}{m_2^2}\left(
\vec{S}_1\cdot\vec{S}_2(\vec{p}_2\cdot{\vec{n}})^2
-\vec{S}_1\cdot\vec{p}_2\vec{S}_2\cdot\vec{n}\vec{p}_2\cdot\vec{n}
-(\vec{S}_1\times\vec{p}_2)\cdot\vec{n}(\vec{S}_2\times\vec{p}_2)\cdot\vec{n}\right)
\right]\nonumber\\
&& +\frac{G^2 (m_1+m_2)}{r^4}\left[7\vec{S}_1\cdot\vec{S}_2-13\vec{S}_1\cdot\vec{n}\vec{S}_2\cdot\vec{n}\right].
\eea
We note that had we implemented the covariant SSC at the level of the action, we would have arrived at the same Hamiltonian appearing here.
Now, we go on to apply the following noncanonical transformations of variables
\bea
\vec{S}_1 &\to& \vec{S}_1 + \frac{p_1^2}{2m_1^2}\vec{S}_1 - \frac{\vec{p}_1\cdot \vec{S}_1}{2m_1^2} \vec{p}_1 \equiv \vec{S}_1 + \frac{\vec{p}_1\times(\vec{S}_1\times \vec{p}_1)}{2m_1^2},\label{eq:scan}\\
\vec{r}_1 &\to& \vec{r}_1 + \frac{\vec{S}_1\times \vec{p}_1}{2m_1^2}-\frac{G}{m_1r^2}\left(\vec{S}_2\times \vec{n}\right) \times \vec{S}_1.\label{eq:rcan}
\eea
These transformations are just the generalization of the mapping between the covariant and NW SSC variables in flat spacetime \cite{Hanson:1974qy}. The transformation of the spin variable in (\ref{eq:scan}) is just similar to that of flat spacetime, whereas for the center of mass coordinate in (\ref{eq:rcan}), a higher order PN transformation, generalized for curved spacetime, is required  \cite{Hergt:2010pa}. Finally, after the application of these transformations, we arrive at the canonical Hamiltonian of \cite{Steinhoff:2007mb} obtained by the ADM Hamiltonian formalism, appearing here in Eq.~(\ref{eq:hss}).

\subsection{Summary of results}

Our results coincide with those of \cite{Steinhoff:2007mb} obtained by the ADM Hamiltonian formalism, and are also  equivalent to those previously calculated in the EFT approach in \cite{Porto:2006bt}, after taking into account the addition appearing in \cite{Porto:2007tt}. In the latter \cite{Porto:2007tt}, the equivalence to the result in \cite{Steinhoff:2007mb} was shown using a canonical transformation relating the results. However, the origin of the discrepancy between the results was not completely clear before. First, there is the addition appearing in \cite{Porto:2007tt}. Our results here in Sec.~\ref{subsec:nw} confirm that to NLO in the spin1-spin2 interaction applying consistently the SSC at the level of the action in the EFT approach is a valid procedure. Here, the additional term  from Eq.~(\ref{eq:lqsscadd}), regarded in \cite{Porto:2007tt} as a spin-orbit effect, is incorporated in the Feynman diagrams, and is thus included as a proper spin1-spin2 Feynman diagram - that of Fig.~3(a). By that we consistently applied the SSC at the level of the action. While in general applying the SSC at the action level is not a valid procedure, it is correct to NLO in the spin1-spin2 interaction due to power counting considerations. 

Second, we noted here in the evaluation of the diagram in Fig.~2(b), where the time derivative term in the spin graviton coupling of Eq.~(\ref{eq:l4}) is present, that there are two ways to evaluate the diagram. The one taken here in Eqs.~(\ref{eq:cd}, \ref{eq:cd1}) is straightforward, but yields acceleration terms in the Lagrangian. Though one may be reluctant to have such terms, calling for the use of the EOM in the Lagrangian, which is known to be incorrect in many examples, the substitution of low order EOM into higher order terms in the level of the Lagrangian is known to be a correct procedure in GR, and it is equivalent to a coordinate transformation. Indeed, we eliminated the acceleration terms by substitution of the LO EOM, and obtained the nonlinear contribution in $G$ in Eq.~(\ref{eq:cd1}), which appeared to be out of place in the one-graviton exchange sector. Hereby, we implicitly had a coordinate transformation made. However, as we noted, there is a second possible way to evaluate this diagram, which avoids acceleration terms, by flipping the time derivative involved between the two particles worldlines. This was the route taken in \cite{Porto:2006bt}, thus requiring the canonical transformation in \cite{Porto:2007tt}, in order to get the result of \cite{Steinhoff:2007mb} in the ADM coordinates. 

To conclude, our results demonstrate exactly that the position, velocity, and spin variables used in the EFT approach relate to the canonical ones via standard coordinate transformations and variable redefinitions, as opposed to the reservations expressed in \cite{Steinhoff:2007mb}. 

\section{Conclusions}

In this paper we made a first application in the PN approximation of the KK reduction over the time dimension, proposed in \cite{Kol:2007bc} within the EFT approach, to calculate the next-to-leading order gravitational spin1-spin2 coupling of a binary of spinning compact objects. The KK decomposition applied to the metric to yield the NRG fields, and the KK reduction applied to the EH action, are shown here to improve both the calculation and the physical interpretation of the interaction substantially. On the more technical level, all vertices simplify, most notably the previously voluminous and complicated three-graviton vertex is exchanged with the simple $\phi F^2$ cubic vertex of the KK reduced action. Moreover, the propagator of the gravitomagnetic vector, which is the most common in this spin interaction, is diagonal, thus facilitating most calculations, and the handling of polarization. On the physical level, the reduced KK action is shown here to be sufficient for the description of the gravitational self-interaction within the stationary approximation to NLO in the spin1-spin2 interaction, with no need for corrections to it. We also find here that similarly to the LO spin1-spin2 interaction mediated exclusively by the gravitomagnetic vector graviton, the NLO spin1-spin2 interaction is also mostly mediated by the vector graviton, with the two first deviations from stationarity, and the leading metric correction term in the spin temporal entries 
associated with this vector graviton.  

\section*{Acknowledgments}

I am grateful to Barak Kol for valuable comments and encouragement. I would like to thank Michael Smolkin for helpful discussions. I thank Rafael Porto for comments on a draft. This research is supported by the Israel Science Foundation Grant No.~607/05, and by the German Israeli Project Cooperation DIP Grant H.52.

\end{document}